\documentclass[12pt]{article}
\usepackage{amssymb,cite,amsmath}
\usepackage[dvips]{color}
\begin{document}
\hoffset=-1truecm
\voffset=-2truecm
\begin{center}
{\large \bf On the solution to the separated
equation in the 3-particle Calogero-Moser problem}\\
\end{center}
\begin{center}
{\bf N.~G.~Inozemtseva\\
University "DUBNA", 141980 Dubna, Moscow Region, Russia}\\
nginozv@mail.ru
\end{center}
\begin{center}
 {\bf J.~Dittrich\\
 Nuclear Physics Institute CAS, 250 68 \v Re\v z, Czech Republic}\\
 dittrich@ujf.cas.cz
\end{center}
\begin{center}
{\bf V.~I.~Inozemtsev\\
BLTP JINR, 141980 Dubna, Moscow region, Russia}\\
inozv@theor.jinr.ru
\end{center}
\abstract{
We propose the exact solution of the equation in separated variable
which appears in the process of constructing solutions
to the quantum Calogero-Moser three-particle problem
with elliptic two-particle potential $g(g-1)\wp(q)$. This solution
is found for special values of coupling constants $g\in {\mathbb Z}, \,
g>1$. It can be used for solving three-paricle CM problem under
appropriate boundary conditions.
}
\section{Introduction}
The problem of finding solutions to quantum integrable finite-dimensional systems
in many cases still remains unsolved. The empirical constructions of such
solutions were important at early stages of the development of the theory
of these systems and lead to many important results being applied to
trigonometric Calogero-Sutherland-Moser systems with the Hamiltonian
of the form
\begin{equation}
\label{1-CMH}
H=\sum_{j=1}^{N}{{p_{j}^2}\over 2}+ g(g-1)\sum_{j>k}^{N}V(q_{j}-q_{k}),
\quad p_{j}=-i{\partial\over{\partial q_{j}}}, 
\end{equation}
for $N$ particles in one dimension
with the two-body potential given by
\begin{equation}
\label{2-sin}
V(q)=a^2 (\sin (aq))^{-2}. 
\end{equation}
The coupling constant $g$ is supposed to be real and chosen as $g>1$.

It turned out \cite{Sut} that the ground-state wave function of the trigonometric
model is of factorized form and the wave functions of all the excitations
can be written as products of this function and multivariable Jack polynomials \cite{Ols}.
These results were also extended to more
complicated cases with interaction terms modified by the introduction of more
general root systems \cite{Ols}.

The quantum elliptic many-particle problem which also has been proven
to be integrable
\cite{Cal,Mos}
is till now quite far from being solved completely.
It has two-particle interaction potential of the form
\begin{equation}
\label{3-wp}
V(q)=\wp(q), 
\end{equation}
where $\wp(q)$ is the Weierstrass elliptic function with two periods
$\omega_{1,2}$ which do not lie on the line in complex plane.
The hermiticity of the Hamiltonian implies
$\omega_{1}\in {\mathbb R},\, i\omega_{2}\in {\mathbb R}$. The trigonometric case (\ref{2-sin})
corresponds to infinite complex period.

In the simplest case of $N=2$, the eigenvalue problem for the Hamiltonian
(\ref{1-CMH}),(\ref{3-wp}) is just the well-known Lam\'{e} equation. For general coupling constants
$g$ its solutions have a branch point at the origin and their expression
through the known transcendental functions is not known.
However, there is an exception: for integer values of $g$, Hermite found that
the solutions are expressed in terms of an exponent and quasiperiodic
Weierstrass $\sigma$ functions \cite{Whi}. The reason for the existence of such a
solution is in fact based on its "good" analytic properties in a complex plane
of the variable $q$: at integer $g$ there is no branch points and
the only singularity is a pole at $q =0$ up to the quasiperiodicity.

This fact inspired the authors in the paper \cite{Dit} to consider the case
of general $N>2$ and $g\in {\mathbb Z}$. It has been proved  that the double
quasiperiodic solutions for many-particle wave functions
are still expressed in terms of the Weierstrass $\sigma$ functions but the
procedure of finding them is rather complicated. They were able to find
it explicitly only for $N=3$, $g=2$. In \cite{Fel}, these solutions have been
presented analytically for arbitrary $N>2$ and $g\geq 2$, also in
overcomplicated form requiring many nontrivial operations to their
explicit writing. As for arbitrary real $g>1$ the solution of the eigenproblem
for the elliptic case was constructed by the perturbation theory in the form
of infinite series \cite{Kom}.

However, there is another approach to finding the solutions
for the dynamics of integrable systems, namely separation of variables.
It is well known in its simple form using purely co-ordinate transformations.
As for elliptic CM systems. simple forms do not work but separation
still take place as it was proposed in \cite{Skl} for 3-particle case at
{\it arbitrary} values of coupling constant $g$.
The separation of variables occurs after transformation corresponding
to a classical canonical transformation of phase space variables
mixing coordinates and momenta. The transformation is realized as an
integral transform of the wave function in the quantum case.
The original two-dimensional problem has been reduced to one-dimensional one
and the process of finding the eigenfunction contains investigation
of the solution to a third order ordinary differential equation \cite{Skl},
\begin{eqnarray}
\nonumber
\psi '''(x)-ih_{1}\psi ''(x)- (h_{2}+3g(g-1)\wp(x))\psi '(x)
\\
\label{4-SklEq}
+[ih_{3}+ig(g-1)h_{1}\wp(x) +g(g-1)(g-2)\wp'(x)]\psi(x)=0,  
\end{eqnarray}
where $h_{1}, h_{2}, h_{3}$ are constants (the values of the integrals
of motion).

The aim of this paper is to find the explicit solutions to equation (\ref{4-SklEq}).
We shall show below that for integer values of $g$, $g\geq 2$ they may
be obtained via the solution to the system of $g$ usual transcendental
equations.

\section{Finding the solution}

It should be noted at first that the coefficients in (\ref{4-SklEq}) are double
periodic functions of $x$. Hence one try to seek the particular solution
as double quasiperiodic function.
The second
observation concerns possible singularities of this solution. Since
$\wp(x)$ has double pole at $x=0$ and it is analytic at the other points of
torus ${\mathbb T}={\mathbb C}/({\mathbb Z}\omega_1 + {\mathbb Z}\omega_2)$,
all the singularities of $\psi(x)$ in ${\mathbb T}$ must be also located at $x=0$.
The assumption
$$
\psi(x)\sim x^{\rho}, \quad \rho<0
$$
at $x\to 0$ gives the result
$$\psi(x)\sim x^{1-g}$$
as the leading singularity, taking the most singular possibility only.
For non-integer $g$, this is a branch point
and there is no simple ansatz to the solution of (\ref{4-SklEq}).

The situation is changed drastically if $g\in {\mathbb Z}$, $g>1$. In
this case, the leading singularity of $\psi(x)$ is a pole of the
order $g-1$ and there are no branch points. Combining this property
with double quasiperiodicity allows one to write down the Hermite-like
ansatz for the possible solution to (\ref{4-SklEq})
\begin{equation}
\label{5-ansatzpsi}
\psi(x)=A \exp(\gamma x){{\prod_{s=1}^{g-1}\sigma(x+\lambda_{s})}\over
{\sigma(x)^{g-1}}}, 
\end{equation}
where $A$ is inessential normalization constant, $\gamma$ and $\{\lambda_{s}\}$
are parameters which have to be determined, and $\sigma(x)$ is the Weierstrass
sigma function.
It is connected with $\wp(x)$
by the relations
$$
{d\over{dx}}\log \sigma(x)=\zeta (x), \quad {d\over{dx}}\zeta(x)=-\wp(x),
$$
where $\zeta(x)$ is the Weierstrass $\zeta$ function with the property
\begin{equation}
\label{6-zetaasympt}
\zeta(x)\to x^{-1} +O(x^3), \quad x\to 0. 
\end{equation}
We assume that all $\lambda_s$ are mutually different for $s=1,\dots,g-1$ and different
from $0$ in ${\mathbb T}$.

 By consecutive differentiations of (\ref{5-ansatzpsi}), one finds
\begin{eqnarray*}
{{\psi'}\over \psi}=\gamma+\sum_{s=1}^{g-1}\zeta(x+\lambda_{s})-(g-1)
\zeta(x),
\\
{{\psi''}\over \psi}=(\gamma+\sum_{s=1}^{g-1}\zeta(x+\lambda_{s})-(g-1)
\zeta(x))^2 -\sum_{s=1}^{g-1}\wp(x+\lambda_{s})+(g-1)\wp(x),
\\
{{\psi'''}\over \psi}=(\gamma+\sum_{s=1}^{g-1}\zeta(x+\lambda_{s})-(g-1)
\zeta(x))^3 +3[\gamma+\sum_{s=1}^{g-1}\zeta(x+\lambda_{s})-(g-1)\zeta(x)]
\\
\times [-\sum_{s=1}^{g-1}\wp(x+\lambda_{s})+(g-1)\wp(x)]
-\sum_{s=1}^{g-1}\wp'(x+\lambda_{s})+(g-1)\wp'(x).
\end{eqnarray*}
Note that all right-hand sides of these equalities are elliptic
functions of the argument $x$. Substitution of these expressions into (\ref{4-SklEq})
 yields
\begin{eqnarray}
\nonumber
B(x)= [\gamma+\sum_{s=1}^{g-1}\zeta(x+\lambda_{s})-(g-1)\zeta(x)]^3
       +3[\gamma+\sum_{s=1}^{g-1}\zeta(x+\lambda_{s})-(g-1)\zeta(x)]
\\
\nonumber
\times [-\sum_{s=1}^{g-1}\wp(x+\lambda_{s})+ (g-1)\wp(x)]-\sum_{s=1}^{g-1}
\wp'(x+\lambda_{s})+(g-1)\wp'(x)
\\
\label{7-Bcompl}
-ih_{1}[(\gamma+\sum_{s=1}^{g-1}\zeta(x+\lambda_{s})-(g-1)\zeta(x))^2
-\sum_{s=1}^{g-1}\wp(x+\lambda_{s})+(g-1)\wp(x)] 
\\
\nonumber
-[h_{2}+3g(g-1)\wp(x)][\gamma+\sum_{s=1}^{g-1}\zeta(x+\lambda_{s})
-(g-1)\zeta(x)]
\\
\nonumber
+ih_{3}+ig(g-1)h_{1}\wp(x)+g(g-1)(g-2)\wp'(x)=0.
\end{eqnarray}
The function $B(x)$ is elliptic and might have poles up to third order
at the points $x=0,\, x=-\lambda_{s} \, (s=1,...g-1)$. However, the direct
inspection  of the Laurent decompositions near these points shows that all
the coefficients at the terms $x^{-3}, x^{-2}$, $(x+\lambda_{s})^{-3},
(x+\lambda_{s})^{-2}$ vanish identically for arbitrary $\gamma$ and
$\{\lambda_{s}\}$. Hence this function can be written in the form
\begin{equation}
\label{8-Bsimple}
B(x)= b_{0}\zeta(x)+\sum_{s=1}^{g-1}b_{s}\zeta(x+\lambda_{s}) +{\rm const},
\end{equation}
where the constant coefficients $b_{0}$, $\{b_{s}\}$ should obey the relation
\begin{equation}
\label{9-polesum}
b_{0}+ \sum_{s=1}^{g-1}b_{s}=0
\end{equation}
(statement (III) of par. 20.12 in \cite{Whi}, e. g.).
The Laurent decomposition of (\ref{7-Bcompl}) near the points $x=-\lambda_{s}$ with the use of (\ref{6-zetaasympt}) allows one to find the coefficients $b_{s}$ explicitly. Due to (\ref{7-Bcompl}), all
of them should vanish. This results in the system of $g-1$ transcendental
equations to the parameters $\gamma$, $\{\lambda_{s}\}$:
\begin{eqnarray}
\nonumber
3[\gamma+\sum_{s\neq k}^{g-1}\zeta(\lambda_{s}-\lambda_{k})+(g-1)\zeta
(\lambda_{k})]^2-3(g-1)^2 \wp(\lambda_{k})-3\sum_{s\neq k}^{g-1}\wp
(\lambda_{s}-\lambda_{k})
\\
\label{10-lambda}
-2ih_{1}[\gamma+(g-1)\zeta(\lambda_{k})+\sum_{s\neq k}\zeta(\lambda_{s}-
\lambda_{k})]-h_{2}=0, \quad k=1,...g-1. 
\end{eqnarray}
It remains only to calculate the constant term in (\ref{8-Bsimple}).
Equivalently, we calculate
$$
\lim_{x\to 0}\left( B(x)-\frac{b_0}{x} \right)
$$
using the Laurent decomposition of (\ref{7-Bcompl}) near the point $x=0$. After long but
straightforward calculations (performed by the
MATHEMATICA$^{\circledR}$
program), one finds
the condition
\begin{eqnarray}
\nonumber
  [\gamma+\sum_{s=1}^{g-1}\zeta(\lambda_{s})]^3 -ih_{1}
  [\gamma+\sum_{s=1}^{g-1}\zeta(\lambda_{s})]^2 -h_{2}
  [\gamma+\sum_{s=1}^{g-1}\zeta(\lambda_{s})]
\\
\label{11-gamma}
+3(2g-3)[\gamma+\sum_{s=1}^{g-1}\zeta(\lambda_{s})]\sum_{s=1}^{g-1}
\wp(\lambda_{s})
+(3g-4)\sum_{s=1}^{g-1}\wp'(\lambda_{s})
\\
\nonumber
- ih_{1}(2g-3)
\sum_{s=1}^{g-1}\wp(\lambda_{s})+ih_{3}
=0.
\end{eqnarray}
The algebraic system (\ref{10-lambda}-\ref{11-gamma}) allows one to determine the parameters
$\gamma$, $\{\lambda_{s}\}$
under which the elliptic function $B$ has no poles and equals zero at one point.
Then $B(x)=0$ due to the Liouville theorem (statement (IV) of par. 20.12 in \cite{Whi}).
The last equation is cubic in $\gamma$. This
corresponds to three linearly independent solutions to the original
equation (\ref{4-SklEq}).

\section{Summary}
Let us summarize our results. We obtained the explicit solutions of the
separated equation (\ref{4-SklEq}) at integer couplings $g$ which, in its turn, gives
the solution to the three-particle quantum Calogero-Moser problem via
the procedure described in \cite{Skl}.
We conjecture that $g$ equations (\ref{10-lambda}-\ref{11-gamma}) determine the $g$ parameters
$\gamma,\, \lambda_s \, (s=1,\dots,g-1)$ in the generic case at the least.
However, it is not clear
whether the solution to the above problem
in the forms known before \cite{Dit,Fel} can be transformed into the forms with
separated variables. As \cite{Dit,Fel,Skl}, we consider in general singular solutions
to the differential equation (\ref{4-SklEq}) leaving aside the right physical boundary conditions
which are even not known here \cite{Skl}.

\vspace{1cm}
\noindent
{\bf Acknowledgement}\\
The work was supported by the
Votruba-Blokhintsev CR-JINR cooperation program in theoretical physics, Czech Science Foundation project 17-01706S and NPI CAS institutional support RVO 61389005.

\end{document}